\documentclass[doublecol]{epl2} 

\usepackage{graphicx}
\usepackage{amssymb}
\usepackage{color}

\title{A fast current-driven instability in relativistic collisionless
  shocks}
\shorttitle{Current-driven instability in relativistic collisionless
  shocks}

\author{Martin Lemoine\inst{1}\thanks{E-mail: \email{lemoine@iap.fr}}
  \and Guy Pelletier\inst{2}\thanks{E-mail:
    \email{guy.pelletier@obs.ujf-grenoble.fr}} \and Laurent
  Gremillet\inst{3}\thanks{E-mail: \email{laurent.gremillet@cea.fr}}
  \and Illya Plotnikov\inst{2}\thanks{E-mail:
    \email{illya.plotnikov@obs.ujf-grenoble.fr}}}
\shortauthor{M. Lemoine \etal}

\institute{
\inst{1} Institut d'Astrophysique de Paris, CNRS, UPMC,
  98 bis boulevard Arago, F-75014 Paris, France\\
\inst{2} UJF-Grenoble 1 / CNRS-INSU, Institut de Plan\'etologie et
  d'Astrophysique de Grenoble (IPAG) UMR 5274,
  F-38041 Grenoble, France\\
\inst{3} CEA, DAM, DIF, F-91297 Arpajon, France 
}
\pacs{52.27.Ny}{Relativistic plasmas}
\pacs{52.35.Qz}{Microinstabilities}
\pacs{52.35.Tc}{Shock waves and discontinuities}

\abstract{We report here on a fast current-driven instability at relativistic
collisionless shocks, triggered by the perpendicular current carried
by the supra-thermal particles as they gyrate around the background
magnetic field in the shock precursor. We show that this instability
grows faster than any other instability studied so far in this
context, and we argue that it is likely to shape the physics of the
shock and of particle acceleration in a broad parameter range.}

\begin{document}
\maketitle

\section{Introduction} 
Fast-growing electromagnetic micro-instabilities are central to
various fields of physics, ranging from astrophysical collisionless
shocks~\cite{1999ApJ...526..697M} to inertial confinement fusion and
high-energy density plasmas~\cite{PhysRevLett.90.155001}. Accordingly,
such instabilities have received ample attention in recent years. Most
notably, the celebrated counterstreaming Weibel/filamentation mode and
its related beam-plasma instability variants (including the oblique
two stream mode~\cite{2004PhRvE..70d6401B}), have been thoroughly
discussed in various scenarios,
e.g.~\cite{2004A&A...428..365W,2006ApJ...647.1250L,2008PhRvL.100t5008B,2010PhRvE..81c6402B,2011MNRAS.417.1148L,2012ApJ...744..182S}. Recent
particle-in-cell (PIC)
simulations~\cite{2008ApJ...681L..93K,2008ApJ...673L..39S,2009ApJ...698L..10N}
have demonstrated that this instability is the agent of mediation of
unmagnetized collisionless shocks, through the build-up of an
electromagnetic barrier on skin depth scales.  Further
simulations~\cite{2008ApJ...682L...5S,2009ApJ...695L.189M,2009ApJ...698.1523S,2009ApJ...693L.127K,2011ApJ...739L..42H,2011ApJ...726...75S,2013ApJ...771...54S}
have shown that the self-generated turbulence provides the scattering
centers required for the development of supra-thermal power-laws, as
anticipated in~\cite{2006ApJ...645L.129L}. Analytical studies have
argued that this same instability may potentially account for the very
efficient pre-heating of the electron population in weakly magnetized
electron-ion shocks, up to near equipartition, which has important
observational consequences in high energy
astrophysics~\cite{2012EL.....9735002G,2013MNRAS.430.1280P}.  Finally,
the Weibel/filamentation instability may also potentially account for
the large magnetization inferred in the external blast of gamma-ray
bursts~\cite{1999ApJ...526..697M}.

In this context, the effective growth rate of the instability is a
prime question, because the fastest mode tends to dominate other
channels of instability (provided it is robust enough, of course). For
relativistic shocks of finite magnetization, this issue becomes acute,
because the rapid advection of the background plasma through the
precursor prevents the growth of slow
modes~\cite{2010MNRAS.402..321L,2011MNRAS.417.1148L}. More
specifically, at large Lorentz factors $\gamma_{\rm
  sh}\,\equiv\,\left(1-\beta_{\rm sh}^2\right)^{-1/2}\,\gg\,1$, and/or
moderate magnetization levels $\sigma$~\footnote{
  $\sigma\,\equiv\,B_{\vert\cal U}^2/\left(4\pi n_{\vert\cal U} m
  c^2\right)$, with $n_{\vert\cal U}$ the proper density of the
  background plasma, $B_{\vert\cal U}$ the magnetic field strength as
  measured in the background plasma rest frame (${\cal U}$).}, even
the Weibel/filamentation mode may not grow on a precursor crossing
timescale. How the shock is structured in such conditions then remains
an open question.

In the present \emph{Letter}, we report on a current-driven
instability which develops generically in a magnetized
counterstreaming beam configuration. As we argue, this instability can
grow faster than the Weibel/filamentation mode in the relativistic
regime and it is the likely agent that shapes the collisionless shock
in a large part of parameter space.  Although we discuss this process
in the framework of a relativistic collisionless pair shock, its range
of applicability is potentially much broader, extending to the field
of beam plasma physics.

The set-up that we have in mind is a beam of ``supra-thermal''
particles gyrating in the magnetic field of the ``background plasma''
in the interpenetration region (e.g., the shock precursor); the
gyrating particles produce an electrical current which is both
transverse to the beam axis and to the background magnetic field; the
compensation of this current by the background plasma then
destabilizes a combination of the extraordinary mode and of the
compressive modes of the plasma, with a growth rate possibly as large
as $\omega_{\rm p}$ in the rest frame of the background plasma. 
In the following, we describe the development of the instability in
the case of a relativistic pair shock; then, we discuss the relevance
of this instability and argue that it can explain recent simulations of
relativistic collisionless shocks.

\section{Current-driven instability}
\subsection{Initial set-up} All throughout, we denote by ${\cal S}$ the
frame in which the shock is at rest, ${\cal U}$ the frame in which the
(far) upstream background plasma is at rest and, without loss of
generality, we assume a shock with perpendicular magnetic field in the
${\cal S-}$frame, with normal incidence.  In the ${\cal S}-$frame,
outside the precursor, the background plasma flows in with
$4-$velocity $u_{x,\infty\vert\cal S} = - \gamma_{\rm sh}\beta_{\rm
  sh}\,<\,0$ and proper density $n_{\vert\cal U}$.  The precursor
contains supra-thermal $e^+-e^-$ (shock-reflected or
shock-accelerated), with mean Lorentz factor $\gamma_{\rm sh}$ and
density $n_{\rm b}=\xi_{\rm b}\gamma_{\rm sh}n_{\vert\cal U}$, which
rotate in the electromagnetic field carried by the background plasma,
i.e., in a magnetic field $\boldsymbol{B}=B_z\,\boldsymbol{z}$ and
convective electric field $\boldsymbol{E}=\beta_{\rm
  sh}B_{z,0}\,\boldsymbol{y}$.  Here, $\xi_{\rm b}\,\sim\,0.1$ denotes
the fraction of incoming shock energy transferred into the
supra-thermal population; the numerical value is inferred from PIC
simulations, e.g.~\cite{2013ApJ...771...54S}.  To zeroth order,
$B_z\,=\,B_{z,0}\,\equiv\,\gamma_{\rm sh}B_{\vert\cal U}$ in terms of
the background rest-frame field $B_{\vert\cal U}$. The gyration of
supra-thermal particles creates a current density $\boldsymbol{j}_{\rm
  b}\,\simeq\, -\xi_{\rm b}\gamma_{\rm sh} n_{\vert\cal U}e
c\,\boldsymbol{y}$.  The size of the precursor is set by the gyration
radius of supra-thermal particles, since these particles execute a
half-gyration orbit in the advected background electromagnetic field
structure in the shock front rest frame ${\cal S}$; these particles
have Lorentz factor $\gamma_{\rm sh}$ and gyrate in
$B_z\,\simeq\,\gamma_{\rm sh}B_{\vert\cal U}$, therefore the size of
this orbit is approximately $c/\omega_{\rm c}$, in terms of
$\omega_{\rm c}=eB_{\vert\cal U}/(mc)$ the upstream cyclotron
frequency.

\begin{figure}
  \includegraphics[bb= 30 230 600 630, width=0.49\textwidth]{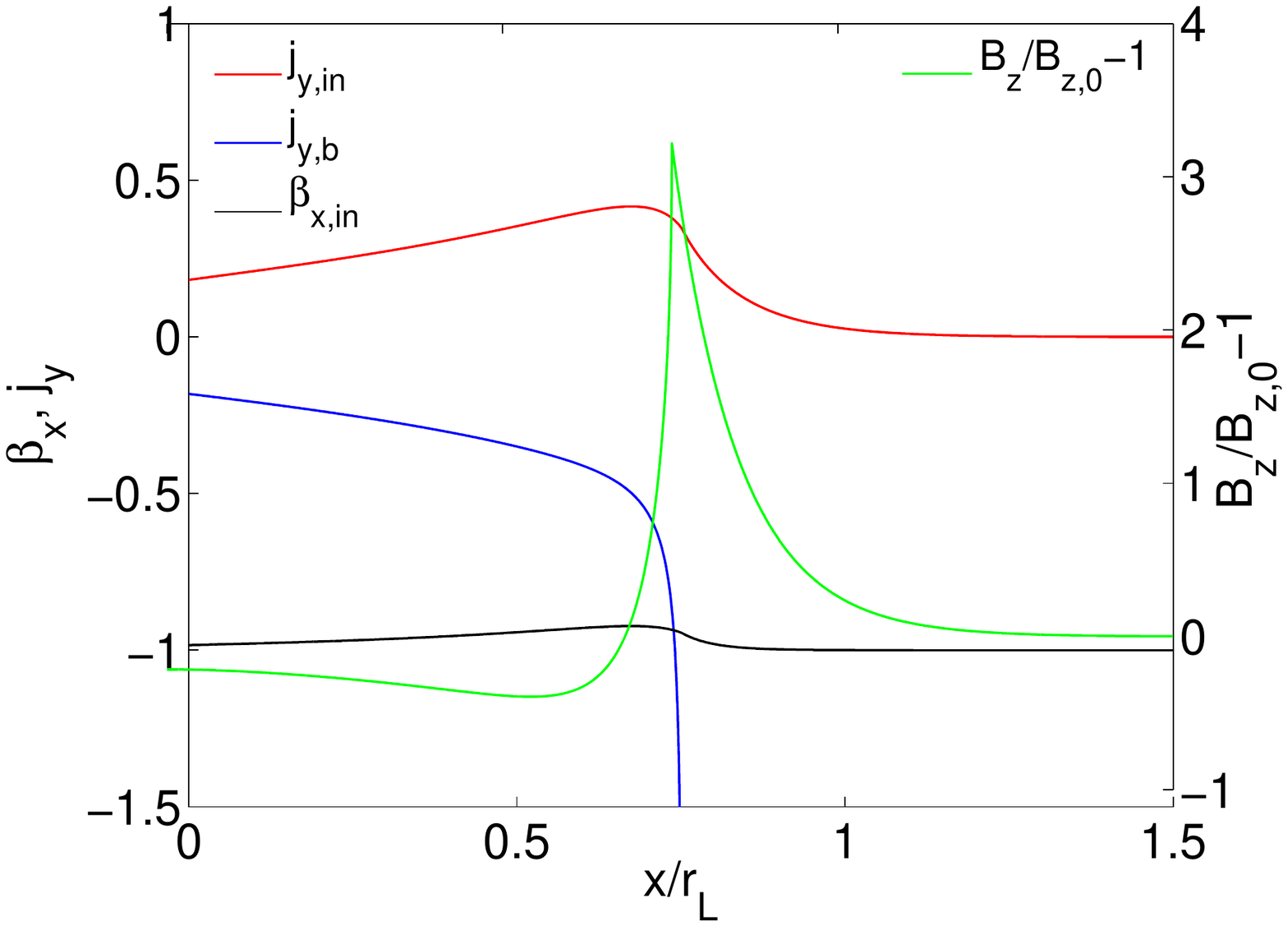}
  \caption{Structure of the precursor as a function of distance to the
    shock, in the shock rest frame, for $\sigma=0.01$, $\xi_{\rm
      b}=0.1$ and $\gamma_{\rm sh}=1000$: profiles of the normalized
    longitudinal velocity, $\beta_x$, transverse current density,
    $j_y$, and magnetic field perturbation, $B_z/B_{z,0}-1$.  The
    indices $\rm in$ and $\rm b$ refer to the background and
    supra-thermal populations. \label{fig:prec} }
\end{figure}

In the ${\mathcal S}$ frame, the incoming background $e^+/e^-$ drift
along $\pm\boldsymbol{y}$ once they enter the precursor, with typical
3-velocity $\beta_y\,\simeq\,\pm\xi_{\rm b}$, in order to compensate
the current $\boldsymbol{j}_{\rm b}$. Their 4-velocity $y-$component
reads: $ u_y \,\simeq\, \pm\gamma_{\rm sh}\xi_{\rm b}c$, hence $\vert
u_y/c\vert\,\gtrsim\,1$ for relativistic shocks, quite possibly $\vert
u_y/c\vert\,\gg\,1$. As discussed in Ref.~\cite{2011MNRAS.417.1148L},
current compensation is achieved to a very high degree. Therefore, in
the following, we treat the supra-thermal particles as rigid and we
study the stability of the system assuming full current
compensation. We will discuss this assumption of a rigid current
further below. Let us already point out that the plasma frequency of
the supra-thermal particle beam is smaller than that of the background
plasma, $\omega_{\rm pb}\,\simeq\,\xi_{\rm b}^{1/2}\omega_{\rm p}$ and
that the growth rate of the current-driven instability peaks at
$\omega_{\rm p}$.

The deflection of the incoming flow along $\boldsymbol{y}$ implies a
deceleration of the flow along $\boldsymbol{x}$: the total ${\mathcal
  S-}$frame 3-velocity remains of order unity,
$\vert\boldsymbol{\beta}\vert\sim 1$, up to corrections of order
$\gamma_{\rm sh}^{-2}$ corresponding to the initial velocity;
therefore, once $\vert\beta_y\vert \,\simeq\,\xi_{\rm b}$, $\beta_x$
has to deviate from unity by quantities of order $\gamma_{\rm
  sh}^{-2}$ or $\xi_{\rm b}^2$, whichever is larger. This leads us to
define a new frame, ${\mathcal R}$, in which there is no bulk motion
of the background plasma along $\boldsymbol{x}$. Assuming in
particular $\gamma_{\rm sh}\xi_{\rm b}\,\gg\,1$, as expected in
ultra-relativistic shocks, one is led to
$\vert\beta_x\vert\,\simeq\,1-\xi_{\rm b}^2/2$, meaning that the
relative Lorentz factor between the shock front rest frame and the
${\cal R-}$frame has dropped from $\gamma_{\rm sh}$ outside the
precursor down to $\gamma_{{\cal R}\vert\rm sh} \,\simeq \, 1/\xi_{\rm
  b}$ inside.  This result holds in a more detailed calculation, which
solves for the spatial profiles of the various fluids and of the
electromagnetic field in a steady state cold fluid approximation; this
model is described in a companion paper~\cite{LPGP13}, but some
salient features are illustrated in Fig.~\ref{fig:prec}.

Note that the upstream plasma is only weakly decelerated as it enters
the precursor, since the $x-$component $u_x$ of its $4-$velocity
decreases by a relative amount $\simeq\,\xi_{\rm b}$. The total
Lorentz factor of each electron and positron components of the
background plasma is approximately conserved to order $\xi_{\rm b}$ as
the components are simultaneously decelerated along $\boldsymbol{x}$
and accelerated along $\pm \boldsymbol{y}$.

Nevertheless, this deceleration along $\boldsymbol{x}$ has important
consequences for the physics of the shock. In particular, the relevant
frame in which to describe the instability as absolute (rather than
convective) becomes ${\cal R}$; in this frame, the relative shock
Lorentz factor with respect to the shock front is $\sim1/\xi_{\rm b}$
and the drift velocity along $y$ in ${\cal R}$ is ultra-relativistic,
$\vert\beta_{y\vert\cal R}\vert\sim1$, because of the Lorentz
invariance of $u_y$.

\subsection{Linear analysis} 
\begin{figure}
\includegraphics[bb=40 150 680 520, width = 0.48\textwidth]{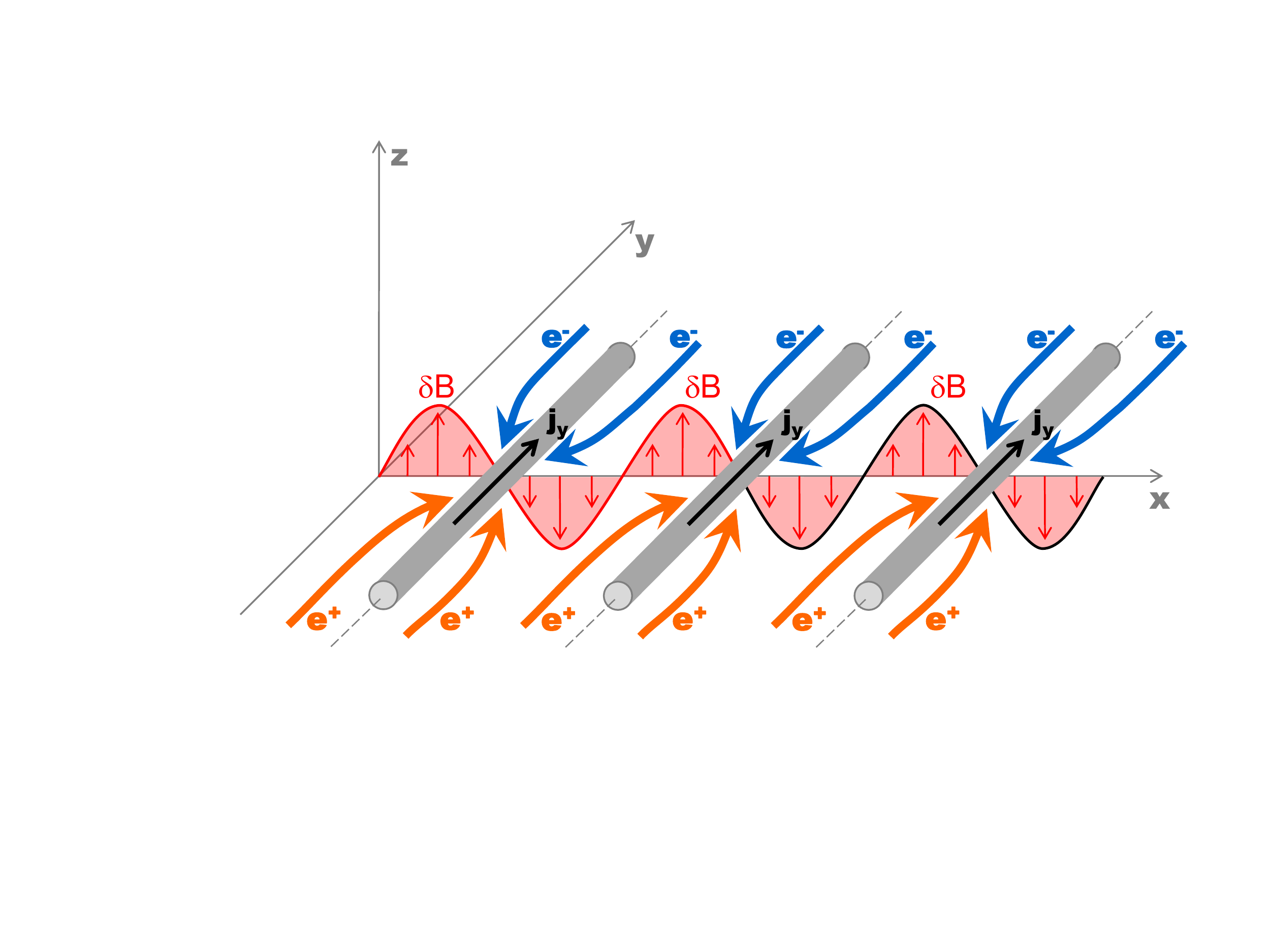}
\caption{ Sketch of the development of the current-driven
  filamentation instability in the background plasma, with a
  perturbation along $\boldsymbol{x}$, in a 1D weakly magnetized
  limit. The external current carried by the supra-thermal particles
  is not shown here; see text for details. \label{fig:sk} }
\end{figure}

The relative drift of electrons and positrons in opposite directions
in the ${\cal R}$ frame gives rise to a current-driven instability,
which can be intuitively understood in the weakly-magnetized
two-dimensional limit $\sigma\,=\,B_{\rm u}^2/\left(4\pi\gamma_{\rm
  sh}^2n_{\vert\cal U}mc^2\right) \,\ll\,\beta_y^2$ for transverse
wavenumbers ($k_y\,=\,0$). Then, it can be described as an analogue of
the Weibel/filamentation instability, with a noticeable difference
related to the opposite nature of the counterstreaming charges: in the
former case, two counterstreaming beams of equal charges are deflected
in opposite directions by a magnetic perturbation, thereby forming a
charge perturbation which couples and feeds back positively onto the
electromagnetic perturbation; this gives rise to filamentation with
currents of opposite polarity in alternating filaments. In the present
situation, the counterstreaming charges are opposite, thus they are
deflected in the same direction at each point and they build a charge
neutral density perturbation. This perturbation also sets up a
perturbed current, which couples and feeds back positively onto the
electromagnetic field. However, the instability now gives birth to
filaments of equal polarity, oriented such as to compensate the
external current induced by the supra-thermal particles (see
Fig.~\ref{fig:sk}).  Such equal-polarity filaments are prone to
coalescence, with potentially important consequences for the
background plasma.

\begin{figure}
\includegraphics[bb=0 70 580 800, width = 0.48\textwidth]{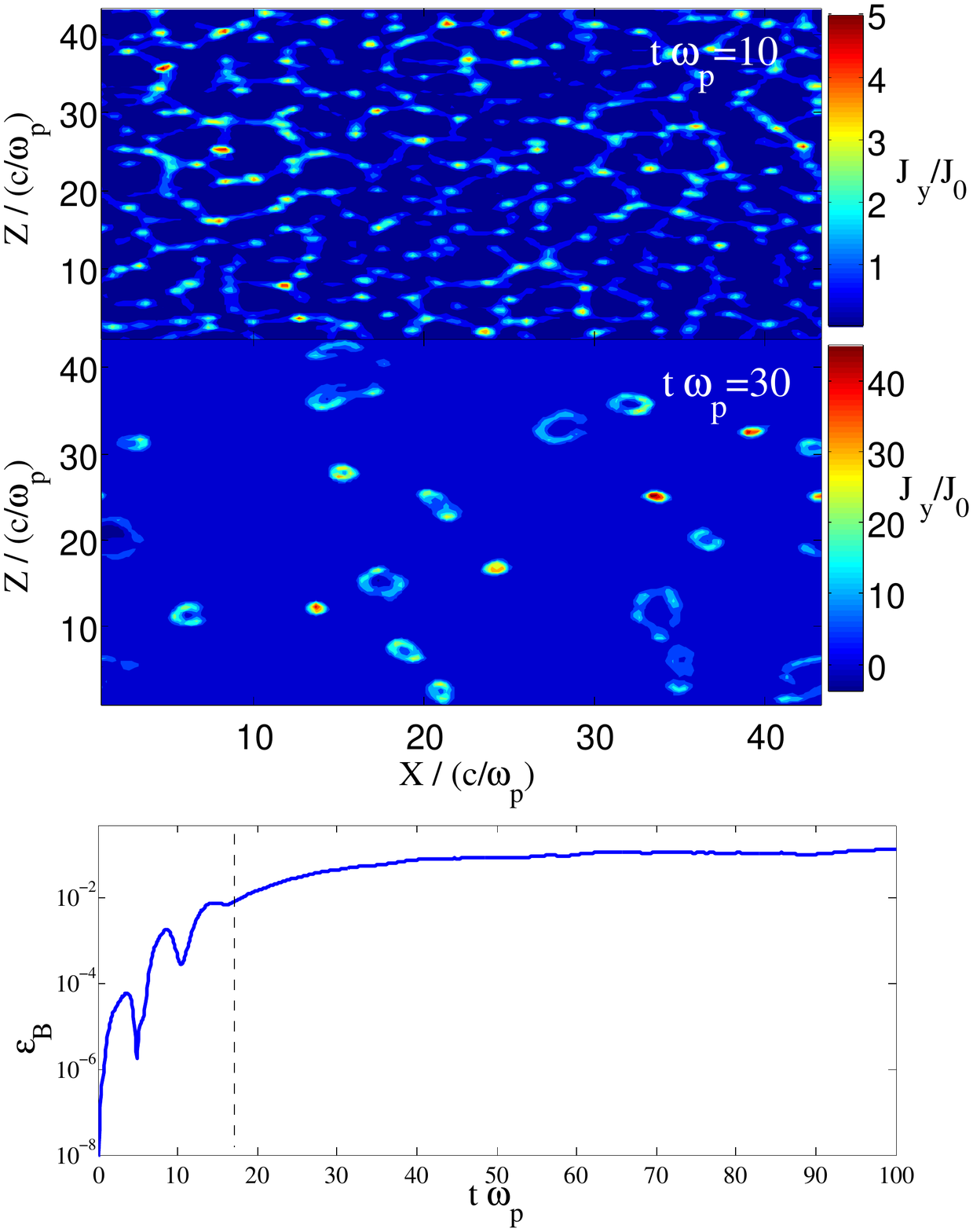}
\caption{2D PIC simulation of the development of the current-driven
  filamentation and of the coalescence phase. See text for
  details.\label{fig:PIC} }
\end{figure}

In order to illustrate the non-linear development of this
current-driven instability in this weakly magnetized limit, we present
in Fig.~\ref{fig:PIC} a snapshot of 2D PIC simulations with the
following set-up: a rigid external current
$\mathbf{j}_0=-enc/\sqrt{2}\, \boldsymbol{y}$ is imposed in the
out-of-plane direction and the simulation box comprises a pair plasma
at zero magnetization; the species drift at velocity
$\beta_y\,\sim\,1/(2\sqrt{2})$, leading to current-driven
filamentation with a growth rate $\Im\omega\,\sim\,0.2\,\omega_{\rm
  p}$, in good agreement with expectations (see thereafter). The
filaments coalesce rapidly: within $100\omega_{\rm p}^{-1}$, there is
only one remaining filament in the $43c/\omega_{\rm p}\times
43c/\omega_{\rm p}$ box. The magnetic field grows exponentially in the
linear phase, then as a power-law in the non-linear phase, starting at
$\sim18\omega_{\rm p}^{-1}$ (indicated by the dashed line), up to a
sub-equipartition value $\epsilon_B \,\sim\, 0.1$. We stress that such
simulations do not aim at reproducing the development of this
current-driven instability in the precursor of a relativistic shock,
since the external current is here held rigid. However, they validate
the general picture that we sketch here. Further simulations in more
general configurations will be reported elsewhere.

In the previous unmagnetized cold plasma limit, one can derive
directly the growth rate from the linear calculation of the
Weibel/filamentation mode: the counter-streaming being symmetric, one
expects an ${\cal R}-$frame growth rate $\Im\omega\,\simeq\,
\vert\beta_{y\vert\cal R}\vert\omega_{{\rm p}\vert \cal R}$, with
$\omega_{{\rm p}\vert \cal R}$ the (relativistic) plasma frequency of
the electrons/positrons in the ${\cal R}$ frame. Here,
$\vert\beta_{y\vert\cal R}\vert\,\sim\,1$ and $\omega_{{\rm p}\vert
  \cal R}\,\simeq\,\omega_{\rm p}$, since the particles have for
apparent density $\sim \gamma_{\rm sh}\xi_{\rm b}n_{\vert\cal U}$ in
the ${\cal R}$ frame and a transverse motion of bulk Lorentz factor
$\sim \gamma_{\rm sh}\xi_{\rm b}$. As a result,
$\Im\omega\,\sim\,\omega_{\rm p}$. To achieve such a fast growth rate
in a highly asymmetrical beam configuration is quite remarkable. This
property directly results from the fact that the external current
imposes the counterstreaming within the background plasma itself,
contrary to the Weibel/filamentation mode which results from the
counterstreaming between the beam and the background plasma.

\begin{figure}
\includegraphics[bb=0 230 580 630, width = 0.48\textwidth]{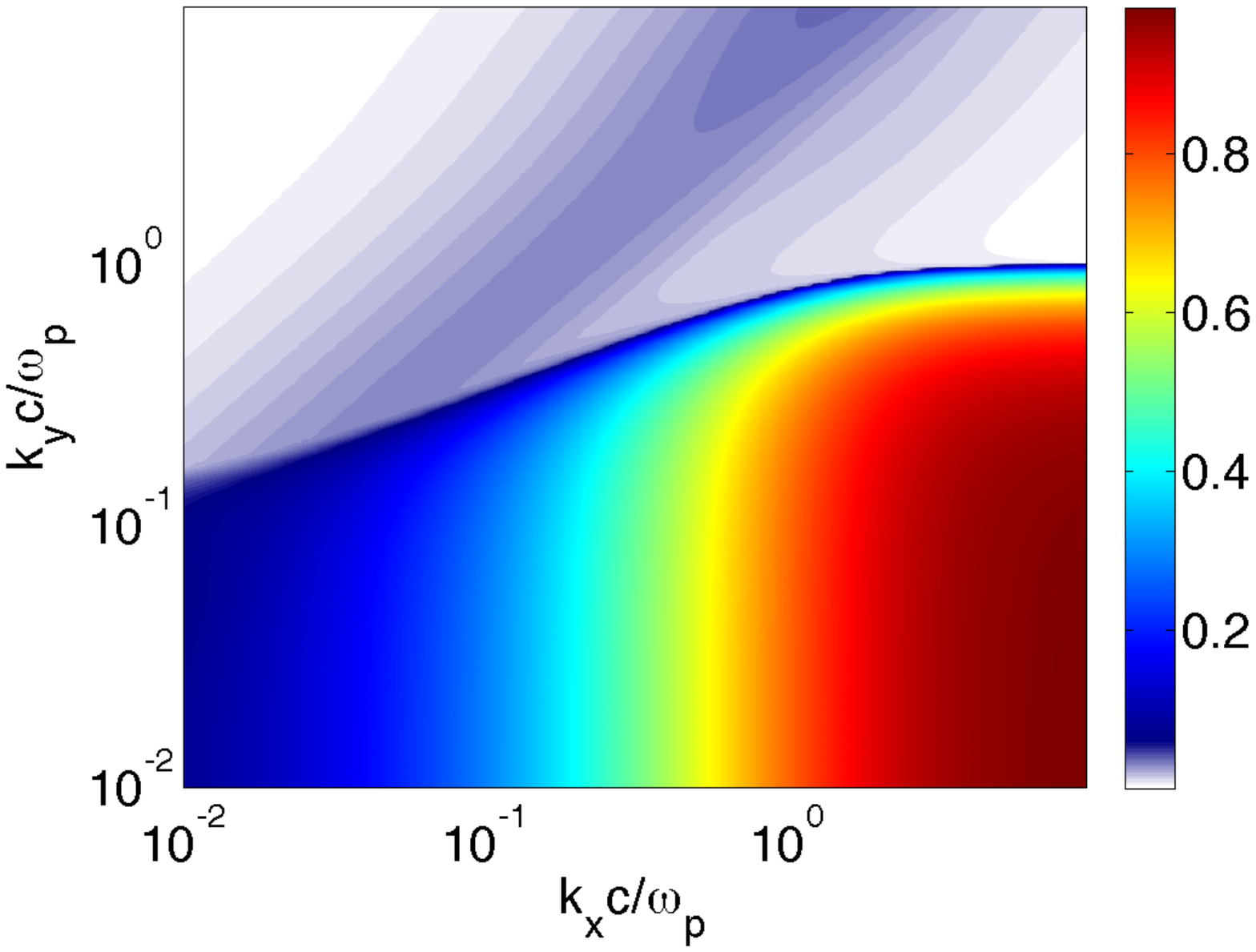}
\caption{Numerical calculation of $\Im\omega/\omega_{\rm p}$ for the
  full dispersion relation, including relativistic effects, assuming
  $k_z\,=\,0$, $\beta_{y\vert\cal R}\,=\,0.99$, $\sigma=10^{-3}$.
\label{fig:gr}}
\end{figure}

One can include finite magnetization effects in a two-fluid model of
the instability: for simplicity, we assume here $\beta_{y\vert\cal
  R}\,=\,\gamma_{\rm sh}\xi_{\rm b}\,\ll\,1$, $k_y=0$ and a cold
background plasma; $\beta_{y\vert\cal R}\,\ll\,1$ means that we can
omit the convective electric field in ${\cal R}$. In
Fig.~\ref{fig:gr}, we provide a numerical solution to the full
dispersion relation of the relativistic two-fluid model described
in~\cite{LPGP13}, for $\beta_{y\vert\cal R}=0.99$, in the 2D plane
$(k_x,k_y)$, for $k_z=0$. This solution indicates that maximum growth
then occurs at values $k_y\,\ll\,k_x$ (similarly, $k_y\,\ll\,k_z$ if
$k_x=0$).  With the present assumptions, one can look for a solution
in which there is no charge density perturbation and $\Delta
u_x=\Delta u_z=0$, with $\Delta u_i \,\equiv\, \left(\delta u_{i+} -
\delta u_{i-}\right)/2$ in terms of the respective velocity
perturbations of the $+/-$ species. The instability involves velocity
fluctuations $\delta u_x$, $\delta u_z$ -- with the definition $\delta
u_i\,\equiv\,\left(\delta u_{i+}+\delta u_{i-}\right)/2$ -- a density
fluctuation $\delta n_{\rm u}\,\equiv\,\left(\delta n_{+,\vert\cal
  U}+\delta n_{-,\vert\cal U}\right)/2$ and an electromagnetic
perturbation characterized by the four-potential component $\delta
A_y$.  Then Maxwell's equations imply in Fourier variables
\begin{equation}
\left(\omega^2-k^2c^2\right)\,\delta A_y\,=\,-4\pi c\delta j_y\ ,\label{eq:m1}
\end{equation}
with 
\begin{equation}
\delta j_y\,=\, 2 e c n_{\vert\cal U}\left(\Delta u_y + \beta_{y\vert\cal R}\frac{\delta
  n_{\vert\cal U}}{n_{\vert\cal U}}\right)\ .
\end{equation}
From the equations of continuity for each species, one derives
\begin{equation}
\frac{\delta n_{\vert\cal U}}{n_{\vert\cal U}}\,=\,\frac{k_xc}{\omega}\delta u_x +
\frac{k_zc}{\omega}\delta u_z\ ,
\end{equation}
so that
\begin{equation}
\left(k^2c^2-\omega^2\right)\delta \hat A_y=\frac{\omega_{\rm
    p}^2}{\omega_{\rm c}}\left(\Delta u_y + \beta_{y\vert\cal R}
\frac{k_xc}{\omega}\delta u_x + \frac{k_zc}{\omega}\delta u_z\right)\ .
\end{equation}
The perturbations $\delta u_x$, $\delta u_z$ and $\Delta u_y$ are
expressed in terms of $\delta \hat A_y\,\equiv\,\delta A_y/B_{\vert\cal U}$
through the equations of motion. One thus ends up with the dispersion
relation
\begin{equation}
\left(P_X - \beta_{y\vert\cal R}^2\omega_{\rm p}^2k^2c^2\right)\omega^2 +
\beta_{y\vert\cal R}^2\omega_{\rm p}^2\omega_{\rm c}^2k_z^2c^2\,=\,0\ ,\label{eq:disp}
\end{equation}
where $k^2\,\equiv\,k_x^2+k_z^2$. $P_X=0$ gives the dispersion
relation of the extraordinary mode: $P_X \,\equiv\,
\omega^4-\left(\omega_{\rm p}^2+\omega_{\rm c}^2+k^2c^2\right)\omega^2
+ \omega_{\rm c}^2k^2c^2$.

The above derivation can be generalized by accounting for thermal
effects through the inclusion of a finite sound velocity. Then the
l.h.s. of Eq.~(\ref{eq:disp}) would be supplemented with a term which
mixes the extraordinary and compressive modes~\cite{LPGP13}. Such
finite temperature effects would quench the instability if $c_{\rm
  s}\,\gg\,\beta_{y\vert\cal R}$; however, one expects
$\beta_{y\vert\cal R}$ close to unity at large $\gamma_{\rm sh}$ while
pre-heating effects in the precursor of pair shocks remain
moderate~\cite{2008ApJ...682L...5S}. The above Eq.~(\ref{eq:disp}) can
be solved in the limit $\sigma\,\ll\,1$ (but not necessarily
$\sigma\,\ll\,\beta_{y\vert\cal R}^2$), leading to a negative root
squared:
\begin{equation}
\label{eq:disp0b}
\omega^2 \,\simeq\, - \left(\beta_{y\vert\cal R}^2 - \sigma\right) \omega_{\rm p}^2
\, 
\end{equation}
for $k c/\omega_{\rm p}\,\lesssim\,1$. The growth rate thus becomes as
high as $\omega_{\rm p}$ as $\beta_{y\vert\cal R}\,\rightarrow\,1$;
this scaling is confirmed by a multi-dimensional study of the
instability, properly accounting for relativistic effects, see
Fig.~\ref{fig:gr} and ~\cite{LPGP13}.

\section{Discussion} \label{sec:disc}
The above current-driven instability grows faster than any of the
other instabilities discussed in the context of relativistic
shocks. In particular, the transverse Weibel/filamentation instability
growth rate is at most $\Im\omega \,\simeq\,\omega_{\rm
  pb}\,=\,\xi_{\rm b}^{1/2}\omega_{\rm p}$ (${\cal R}-$frame). One
thus expects that the current-driven instability represents a key
instability in collisionless shock physics and indeed, this can be
argued as follows.

At relativistic collisionless shock waves of finite magnetization, the
restricted timescale on which a plasma element crosses the precursor,
$t_{\times\vert\cal U}\,\simeq\, (\gamma_{\rm sh}\omega_{\rm c})^{-1}$
(written here in the background plasma rest frame) strongly constrains
which instability can or cannot
grow~\cite{2010MNRAS.402..321L,2011MNRAS.417.1148L}: in particular,
neglecting any deceleration of the flow, the Weibel/filamentation
instability cannot grow in the precursor if
$\sigma\,\gtrsim\,\gamma_{\rm sh}^{-2}\xi_{\rm b}$ -- a very low value
if $\gamma_{\rm sh}\,\gg\,1$ -- as the growth timescale then becomes
larger than $t_{\times\vert\cal U}$. In contrast, the present
current-driven filamentation can grow in a much broader range of
parameter space, due to the deceleration imparted to the background
plasma: at large values of the shock Lorentz factor, i.e. $\gamma_{\rm
  sh}\xi_{\rm b}\,\gg\,1$, the relevant frame in which to discuss the
instabilities as absolute has become ${\cal R}$, which moves with a
$\gamma_{\rm sh}-$independent Lorentz factor $\gamma _{{\cal
    R}\vert\rm sh}\,\sim\,1/\xi_{\rm b}$ relatively to the shock
front.  In this ${\cal R-}$frame, the precursor crossing timescale
becomes $t_{\times\vert\cal R}\,\simeq\,1/(\gamma_{{\cal R}\vert\rm
  sh}\omega_{\rm c})$; hence, the current-driven instability can grow
whenever $\Im\omega\, t_{\times\vert{\cal R}}\,\gtrsim\,1$, or
\begin{equation}
\sigma\,\lesssim\,\xi_{\rm b}^2\ .\label{eq:cdf}
\end{equation}
For $\sigma\,\lesssim\,10^{-2}$ (assuming $\xi_{\rm b}\,\sim\,0.1$),
this implies that the current-driven filamentation can grow at
\emph{any} value of the shock Lorentz factor.  Therefore, at large
Lorentz factors and/or moderate magnetization levels, in particular
$\gamma_{\rm sh}^2\sigma\xi_{\rm b}^{-1}\,\gtrsim\,1$ such that the
Weibel/filamentation instability cannot grow, this current-driven
instability is expected to shape the physics of the precursor.

\begin{figure}
\includegraphics[bb=60 20 650 400, width = 0.48\textwidth]{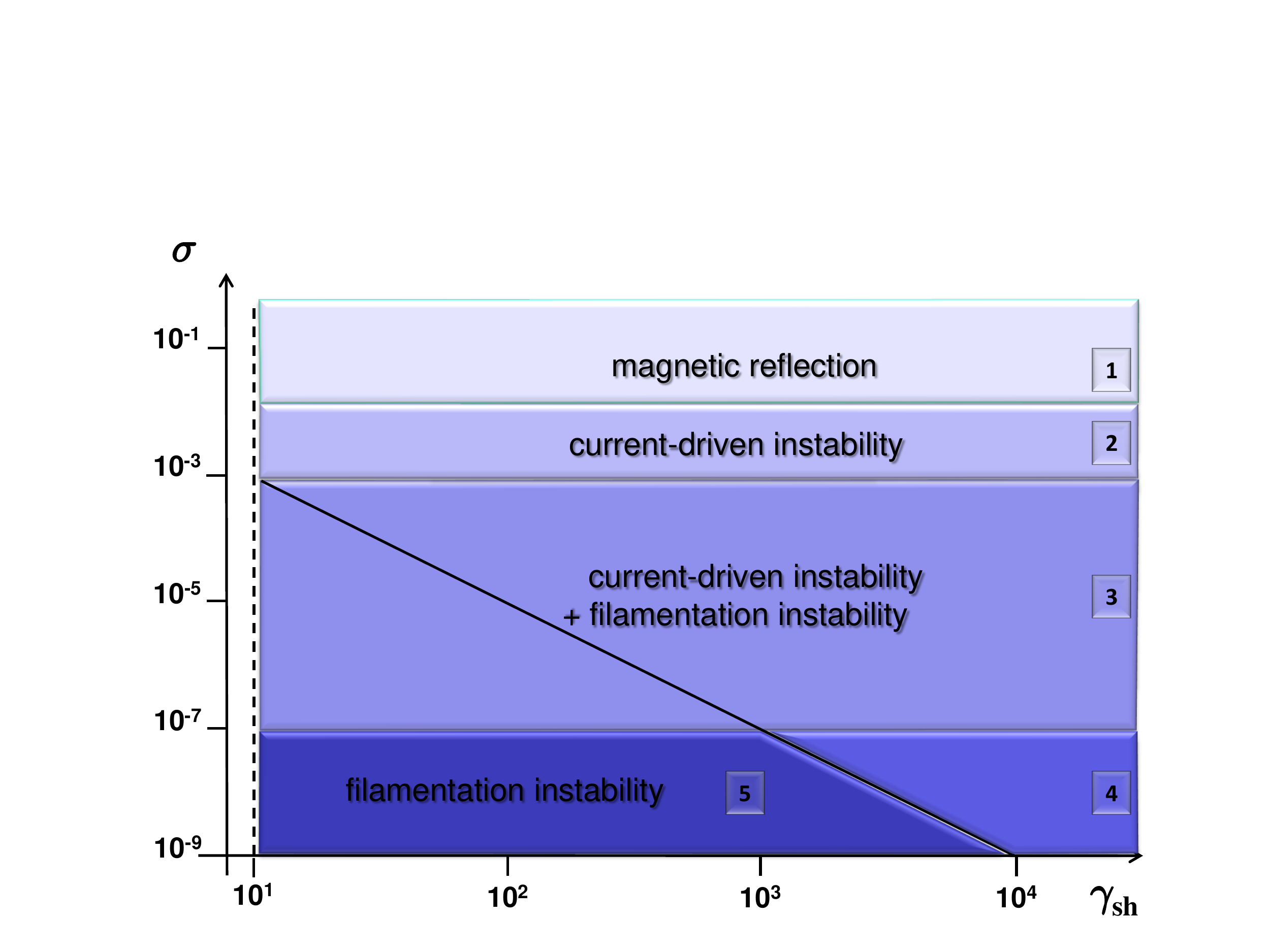}
\caption{Phase diagram of relativistic $e^+e^-$ collisionless shocks
  in the plane $(\gamma_{\rm sh},\sigma)$. Our calculations apply to
  the highly relativistic regime, to the right of the dashed line
  ($\gamma_{\rm sh}\,\gtrsim\,10$); this figure assumes $\xi_{\rm
    b}=0.1$. The various regions indicate the zones of influence of
  various instabilities and processes. Region 1: the shock transition
  is initiated by magnetic reflection; region 2: the current-driven
  instability dominates, for $\sigma\,\lesssim\,\xi_{\rm
    b}^2\,=\,10^{-2}$; region 3: for $\sigma\,\lesssim\,\xi_{\rm
    b}^3\,=\,10^{-3}$, both the current-driven and the filamentation
  instability can grow, thanks to the deceleration imparted by current
  compensation; region 5: at low values $\sigma\,\lesssim\,10^{-7}$
  (see text for details), micro-turbulence weakens the perpendicular
  current, the filamentation instability likely dominates the physics;
  in region 4, microturbulence does weaken the perpendicular current,
  but the filamentation mode could not grow (above the diagonal line
  $\gamma_{\rm sh}^2\sigma\xi_{\rm b}^{-1}\,=\,1$) without the
  deceleration imparted by current compensation. See the text for
  further details.
 \label{fig:phase}}
\end{figure}

The domain of influence of this instability in a phase diagram of
relativistic collisionless shocks is illustrated in
Fig.~\ref{fig:phase}. At high magnetization, the shock is known to be
mediated by magnetic reflection and the synchrotron maser instability,
e.g. ~\cite{1988PhFl...31..839A,1991PhFlB...3..818H,1992ApJ...390..454H,1992ApJ...391...73G}. At
very low magnetizations, the Weibel/filamentation mode can structure
the shock, as demonstrated by PIC simulations in the unmagnetized
limit, e.g.~\cite{2008ApJ...682L...5S}. However, this same
filamentation instability cannot grow above the diagonal line
corresponding to $\gamma_{\rm sh}^2\sigma\xi_{\rm
  b}^{-1}\,\gtrsim\,1$, as discussed above, in the absence of
deceleration imparted by current compensation. In contrast, the
current-driven instability can grow for any value of the shock Lorentz
factor, provided $\sigma\,\lesssim\,10^{-2}$; this determines its zone
of influence illustrated in Fig.~\ref{fig:phase}.

Of course, once the background plasma has been decelerated through
current compensation, the criterion for the growth of the
Weibel/filamentation instability is itself modified; repeating the
above arguments in ${\cal R}$, one finds $\sigma \,\lesssim\,\xi_{\rm
  b}^3$, which remains more stringent than for the current-driven
filamentation. The two instabilities can then operate in conjunction
in the precursor of relativistic shocks for
$\sigma\,\lesssim\,\xi_{\rm b}^3$, thanks to the deceleration imparted
by current compensation; however, in this region, the
Weibel/filamentation instability still grows more slowly than the
current-driven instability in the ${\cal R}$ frame. Which instability
prevails in this zone should be determined by PIC simulations. We
comment further on this issue in the following.

The micro-turbulence not only builds up the shock at low
magnetization, it also provides the needed source of scattering for
the relativistic Fermi process. The analysis of particle trajectories
actually leads to the following {\it sine qua non} condition for a
proper relativistic Fermi process, see
Ref.~\cite{2010MNRAS.402..321L}:
\begin{equation}
\sigma\,\lesssim\,\epsilon_{B,\rm d}^2\ ,\label{eq:F}
\end{equation}
where $\epsilon_{B,\rm d}\,\sim\,10^{-2}$--$10^{-1}$ represents the
typical value of $\epsilon_B$ downstream of the shock, assuming a
micro-turbulence on scales $c/\omega_{\rm p}$. In this regard, the
current-driven instability also plays a key role in triggering
relativistic Fermi acceleration: by building up the micro-turbulence
for any value of the shock Lorentz factor, it gives rise to the Fermi
process in a large fraction of parameter space, whereas the standard
Weibel/filamentation instability would be limited to the region
$\sigma\,\lesssim\,\gamma_{\rm sh}^{-2}\xi_{\rm b}$. In this respect,
we note that the transverse kinetic energy flux is a factor
$\left\vert u_y/u_{x,\infty\vert\cal S}\right\vert\,\simeq\,\xi_{\rm
  b}$ of the incoming kinetic energy flux along $\boldsymbol{x}$,
which implies that the waves excited through the current-driven
instability may potentially grow up to a fraction
$\epsilon_B\,\sim\,\xi_{\rm b}$, enough to mediate the shock.

The picture that we are drawing here seems supported by recent PIC
simulations. In particular, one expects to observe a precursor of
scale height $\sim\,c/\omega_{\rm c}$ if the current-driven
instability develops, since $c/\omega_{\rm c}$ sets the scale over
which the perpendicular current is generated; in contrast, the scale
height of the precursor in Weibel/filamentation mediated shocks should
be independent of that scale. The simulations reported in
~\cite{2013ApJ...771...54S} precisely find a precursor with a scale
height $\simeq\,2c/\omega_{\rm c}$ (as evaluated from their figure~7);
these simulations use $\gamma_{\rm sh}=21$ and
$10^{-4}\,\lesssim\,\sigma\,\lesssim\,10^{-3}$, values for which the
Weibel/filamentation alone could not be excited, contrary to the
current-driven instability. Moreover, the magnetic field appears
structured in sheets parallel to the $x-y$ plane in these simulations
rather than in filaments oriented along $\boldsymbol{x}$ (see their
fig.~5), as would be expected for a standard Weibel/filamentation
instability. Although our present linear analysis cannot predict the
topology of the structures grown from the current-driven instability,
we find that maximal growth occurs for $k_y\,\ll\,k_z$, assuming a
vanishing $k_x$, suggesting that the $z-$dependence is an important
factor; one should of course investigate these issues with dedicated
PIC simulations for a proper comparison to the above simulations.

Furthermore, the simulations of Ref.~\cite{2013ApJ...771...54S} report
no dependence on the shock Lorentz factor, whereas a clear signature
should be seen for the Weibel/filamentation mode, when the line
$\gamma_{\rm sh}^2\sigma\xi_{\rm b}^{-1}=1$ is crossed. In the present
picture, this independence is understood to result from the
deceleration to $\gamma_{{\cal R}\vert\rm sh}\,\sim\,1/\xi_{\rm b}$
imparted by current compensation, the precursor playing the role of a
buffer.

As the magnetization decreases, the micro-turbulence in the shock
precursor plays an increasingly important role in scattering the
supra-thermal particles. If this micro-turbulent scattering comes to
dominate the transport of supra-thermal particles, it will randomize
the tangential components of their velocities, hence it will randomize
the perpendicular current itself; conversely, if coherent gyration
around the background field prevails, the supra-thermal particle
current can indeed be considered as rigid. This allows to derive a
critical value of $\sigma$, below which the back-reaction of the
micro-turbulence becomes important, and above which our assumption of
a rigid external current is valid, as follows. During a time interval
$t_{\cal R}$, the particle acquires a rms deflection
$\delta\theta\,\sim\,\left(\nu_{\rm s\vert\cal R}t_{\cal
  R}\right)^{1/2}$ through micro-turbulent scattering; in order for
the particle to gyrate back to the shock front, it needs to be
deflected away from the shock normal by an angle $\delta \theta_{\cal
  R}\,\sim\, 1/\gamma_{\cal R\vert\rm sh}\,\sim\,\xi_{\rm b}$ in the
${\cal R}$ frame, see
e.g.~\cite{2001MNRAS.328..393A,2006ApJ...651..979M,2013MNRAS.430.1280P};
therefore, micro-turbulent scattering dominates over coherent gyration
if $\delta \theta\,\gtrsim\,\delta\theta_{\cal R}$ for $t_{\cal
  R}\,\sim\,\delta\theta_{\cal R}/\omega_{\rm c\vert\cal R}$, the
latter corresponding to the time taken to achieve a deflection of
$\delta\theta_{\cal R}$ through coherent gyration. Using the fact that
supra-thermal particles have typical Lorentz factor $\gamma_{\rm
  sh}/\xi_{\rm b}$ in the ${\cal R}$ frame, assuming that the
micro-turbulence peaks on skin depth scales in this frame, one finds
that coherent gyration remains a good approximation if
$\sigma\,\gtrsim\, \xi_{\rm b}^2\epsilon_{B,\rm u}^2$,
$\epsilon_{B,\rm u}$ characterizing the strength of the
micro-turbulence in the shock precursor. Therefore, for
$\sigma\,\gtrsim\,10^{-7}$, coherent gyration is expected to prevail
even up to the large values of $\epsilon_{B,\rm u}$ reached close to
the shock front, supporting our assumption of a rigid external
current. At lower values of $\sigma$, one should still expect the
current-driven instability to develop at the tip of the precursor,
where $\epsilon_{B,\rm u}$ is small; how the whole precursor is
structured in this limit remains open for further study.

This current-driven instability thus modifies the paradigm according
to which weakly magnetized ($\sigma\,\ll\,1$) shock waves are mediated
by the Weibel/filamentation mode. Actually, the only region in
parameter space where one can expect the Weibel/filamentation mode to
grow in the absence of current-driven filamentation is
$\sigma\,\lesssim\,\gamma_{\rm sh}^{-2}\xi_{\rm b}$.

The existence of this current-driven instability opens new avenues of
research. In the field of collisionless shock physics, one must
notably explore the exact interplay between this instability and the
Weibel/filamentation mode in the region in which they can both grow,
using dedicated PIC simulations. One must also understand how this
instability shapes the magnetic field structure of the precursor,
since the properties of the turbulence directly impact the
acceleration process, e.g.~\cite{2013MNRAS.430.1280P}. This
instability should also play an important role at the termination
shock of pulsar winds, for which one expects $\sigma
\,\lesssim\,10^{-2}$ and $\gamma_{\rm sh}\,\gg\,100$; which role and
whether it can explain the remarkable (and unexpected) acceleration
efficiency of such shocks~\cite{2012SSRv..173..341A}, remains to be
understood.

\bibliography{shock}

\end{document}